\documentclass[12pt]{spieman}  
\usepackage{amsmath,amsfonts,amssymb}
\usepackage{graphicx}
\usepackage{setspace}
\usepackage{tocloft}
\usepackage{lineno}
\usepackage{xcolor}
\usepackage{color, soul}
\sethlcolor{yellow}
\usepackage{hyperref}
\soulregister\cite{7}
\soulregister\ref{7}
\soulregister\pageref{7}
\linenumbers

\title{Voltage-to-temperature calibration of the High Altitude THz Solar telescope acquisition system}

\author[a]{Gedeane G.S. Kenshima}
\author[a]{Daniel R. Sousa}
\author[b,e]{Tiago Giorgetti}
\author[a,c,*]{Paulo J. A. Simões}
\author[a,d]{C. Guillermo Giménez de Castro}
\affil[a]{Center for Radio Astronomy and Astrophysics Mackenzie, Engineering School, Mackenzie Presbyterian University, São Paulo, SP, Brasil}
\affil[b]{Giorgetti Engenharia, São Paulo, SP, Brasil}
\affil[c]{School of Physics and Astronomy - University of Glasgow, Glasgow, UK}
\affil[d]{Instituto de Astronomía y Física del Espacio, Ciudad Autónoma de Buenos Aires, Argentina}
\affil[e]{Escola Politécnica da Universidade de São Paulo - Poli/USP, São Paulo, SP, Brasil}

\cftpagenumbersoff{figure}
\cftpagenumbersoff{table} 
\begin{document} 
\nolinenumbers
\maketitle

\begin{abstract}
The THz range has been under-explored for solar astronomy, mainly due to technological limitations. Only recently, a few new telescopes, such as the High Altitude Terahertz Solar (HATS) photometer, have been monitoring the solar activity in this range of the spectrum. This paper presents the experimental characterization and voltage-to-temperature calibration of the HATS acquisition system. HATS uses a Golay cell for capturing the incoming radiation, modulated by a 20~Hz fork chopper. The amplitude of the signal is then obtained at predetermined time intervals by applying a windowing function and an FFT to the signal. To characterize this acquisition system, a blackbody calibrator, with temperatures varying from 100 to 500$^{\circ}$C ($373.15$ K to $648.15$ K), was used as a THz source, and two signal recovery methodologies were compared: sinusoidal curve fitting and FFT combined with six windowing functions (rectangular, Hamming, Hann, Barlett, Blackman, and Flat Top). Our quantitative results demonstrate high linearity in the system's response over the input source's temperature range, with the Hamming window achieving the highest precision, yielding a Root Mean Square Error (RMSE) of 15.48 and a calibration temperature-to-voltage factor of $10.32 \pm 0.13$ mV/K. In contrast, the Bartlett window presented the highest error (RMSE 16.02). The choice of windowing function had only a minor effect on the calibration factors, which ranged from 10.32 to 10.43 mV/K, all within the experimental uncertainties. Beyond solar photometry, the signal modulation and windowing concepts established here are highly applicable to other fields requiring high-sensitivity thermal detection, such as industrial pyrometry for high-temperature manufacturing, environmental monitoring of atmospheric water vapor, and the development of medical imaging systems based on Terahertz radiation for non-invasive tissue analysis.
\end{abstract}


{\noindent \footnotesize\textbf{*}Paulo J. A. Simões,  \linkable{paulo@craam.mackenzie.br} }

\begin{spacing}{2}   

\section{Introduction}
\label{sect:intro}  
Solar flares are phenomena characterized by the sudden release of magnetic energy in the solar atmosphere, resulting in emissions that span the entire electromagnetic spectrum \cite{shibata2011, Fletcher2011SSRv..159...19F}. While radio and X-ray observations are well-established, the mid-infrared and Terahertz (THz) domains represent an essential frontier for understanding the chromospheric response during magnetic reconnection processes \cite{Kaufmann2013,penn2016,Lopez2022A&A...657A..51L}. Recent studies emphasize that the infrared continuum emission in solar flares is predominantly of thermal origin, arising from dense layers of the solar atmosphere where local thermodynamic equilibrium is maintained \cite{Trottet2015, Simoes2017A&A...605A.125S,Guigue2018SpWea..16.1261G,Simoes2024,Yang2025ApJ...988L..56Y,Rojas2026arXiv260506344R}.

\begin{figure}
    \begin{center}
    \begin{tabular}{c}
    \includegraphics[height=5.5cm]{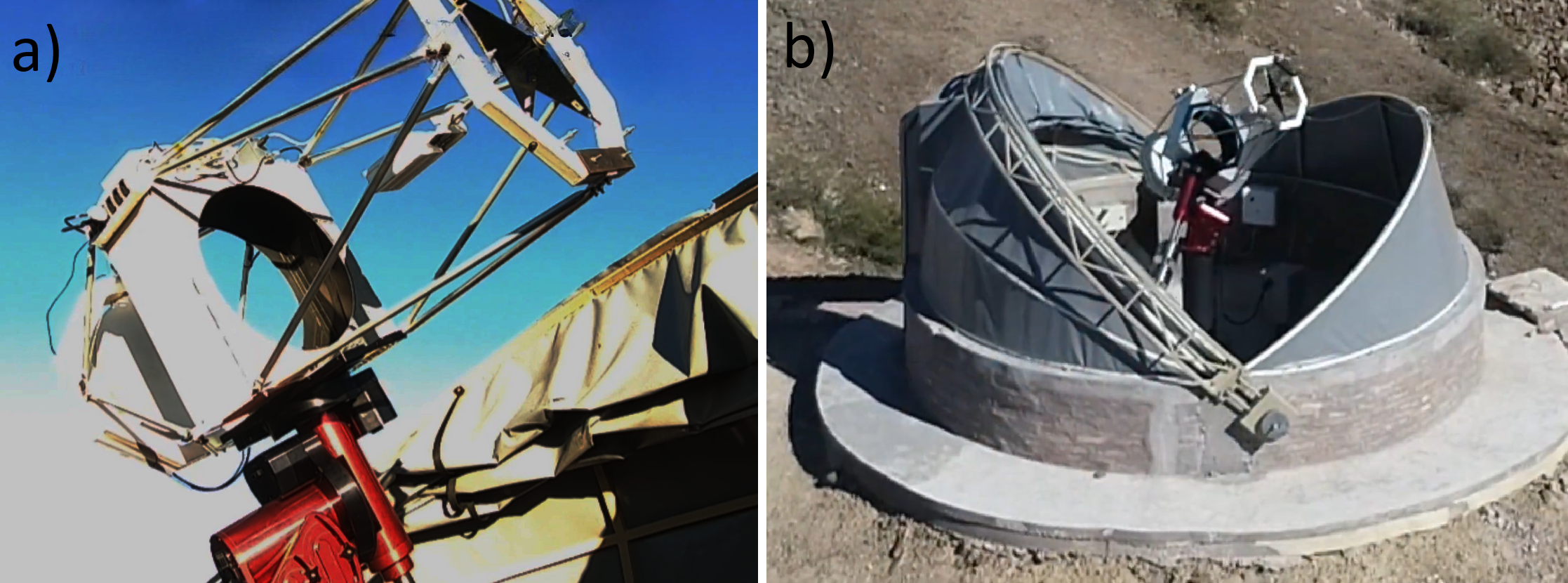}
    \end{tabular}
    \end{center}
    \caption{HATS telescope, operated in OAFA in the Argentinean Andes, with views from (a) inside and (b) outside the dome.\label{fig:HATS}}

\end{figure}

Ground-based observations at THz frequencies face significant technical challenges, primarily due to strong atmospheric absorption by water vapor and the requirement for detectors with high thermal sensitivity \cite{Penn2014LRSP...11....2P}. The High Altitude THz Solar photometer \cite{Castro2020} (HATS, Figure \ref{fig:HATS}), installed at the Observatorio Astronómico Félix Aguilar (OAFA), was designed to mitigate these limitations by operating at an altitude that allows for the detection of rapid solar brightness variations at 15~THz \cite{Castro2020}. 

A precise correlation between the detector's output voltage and the source's brightness temperature is necessary to produce scientifically interpretable data. HATS optics and acquisition system have been previously detailed \cite{Castro2020}, and a computational analysis of windowing functions for transient signal recovery and noise reduction via the Fast Fourier Transform (FFT) was evaluated \cite{Kenshima2026}. The present study focuses specifically on the experimental voltage-to-temperature calibration required to convert raw data into physical brightness temperature.

\section{Instrumentation and calibration setup}

The core of the HATS detection system is based on Golay cells, which are detectors known for their flat spectral response over a wide frequency range \cite{Golay10.1063/1.1740948,fernandes2011}. The use of these detectors in solar photometry requires input signal modulation via a mechanical chopper, allowing the electronic system to process only the alternating component of the radiative signal \cite{fernandes2011,Yingxin6380076}. This experimental configuration enables the detection of subtle thermal fluxes, which are essential for characterizing the chromospheric temperature both at quiet times and during impulsive events \cite{kudaka2015, Miteva2016}.

The absolute calibration of the instrument is grounded in blackbody radiation laws, specifically employing the Rayleigh-Jeans approximation for the low-frequency regime relative to the thermal emission peak \cite{1986rpa..book.....R, stix2004}. In the proposed experimental setup, a high-precision blackbody calibrator serves as the reference source, with its temperature varied to map the acquisition system's response. This procedure aims to determine the transfer coefficients that convert digitized values into physical units of irradiance, enabling direct comparison with theoretical models of radiative transport in the solar atmosphere \cite{1986rpa..book.....R, Simoes2017A&A...605A.125S}.

Although the quiet Sun brightness temperature at THz frequencies is approximately $5000\text{ K}$, the effective temperature variation reaching the primary detector is significantly lower due to atmospheric attenuation and the overall telescope beam efficiency ($\eta_{MB} \approx 0.11$, accounting for optics, chopper, and spatial beam filling factors \cite{GimenezdeCastroetal:2026}). Consequently, the radiation power incident on the Golay cell during solar observations corresponds to an equivalent brightness temperature deflection of only a few hundred Kelvin (around $500\text{ K}$). Therefore, the blackbody calibration temperature range of $100^\circ\text{C}$ to $375^\circ\text{C}$ ($373.15\text{ K}$ to $648.15\text{ K}$) appropriately matches the linear dynamic range expected at the sensor input under real observational conditions.

\section{Methodology}

Data were obtained in bench tests using a Golay cell model, Tydex GC-1P, and an Omega BB-4A blackbody calibrator with variable temperature. We recorded, using the Golay cell, the THz signals from the blackbody source, varying its temperature from $T_{min}= 100^{\circ}$C (373.15 K) ~to~$T_{max}=375^{\circ}$C (648.15 K) in $50^{\circ}$C intervals. Each temperature reading lasts 200 ms, and the interval between readings is 15 minutes, the time required for the blackbody reference source to stabilize. As the measurements varied, both increasing and decreasing from $T_{min}$ to $T_{max}$, the same blackbody temperature was recorded multiple times. To obtain a single signal for each blackbody temperature, the average signal for that temperature was used.

For each blackbody temperature, we obtained the Golay output signal voltage amplitude and fitted a linear regression 
\begin{equation}
    U = aT + b \ ,
    \label{eq:gain}
\end{equation}
where $T$ is the blackbody temperature, $U$ is the output voltage, $a \ [\mathrm{mV\ K^{-1}}]$  is the \textit{gain}, and $b$ [mV] is an offset. We applied two different approaches to get $U$: 1) \textbf{Windowed FFT}: the amplitude at 20 Hz in the frequency domain of the signal, and 2) \textbf{Sine}: the amplitude of a sine function model adjusted to the signal. The coefficients obtained by the FFT  method were compared with those obtained using the sine function method.

\subsection{Window functions}

We use the windowed Fourier Transform to extract the signal amplitude at 20 Hz. Windowing is the operation of multiplying a finite-duration digital signal $x(n)$ by a window function $\psi(n)$ prior to applying the FFT, 
\begin{equation}
    X_w(k) = \sum_{n=0}^{N-1}x(n)\psi(n) e^{-j 2 \pi nk/N} \ , 
    \label{eq:wfft}
\end{equation}
where $X_w(k)$ is the windowed frequency-domain discrete Fourier Transform (DFT) output for the \(k\)-th frequency, $N$ is the total number of samples in the discrete signal (from $0$ to $N-1$), and $x(n)$ is the input signal value at the $n$-th sample. This method is widely used in Digital Signal Processing (DSP) to reduce spectral leakage, i.e., the dispersion of energy across different frequencies when applying FFT to a finite time series \cite{Jwo2021}. 

In the literature, dozens of window functions have been reported \cite{prabhu2014}. For the present work, we selected six of them: rectangular (or boxcar), Hamming, Hann, Bartlett, Blackman, and Flat-top \cite{Jwo2021,prabhu2014,Bojkovic2017}. Here, we employ the windows available in \texttt{scipy.signal.windows}\cite{scipy_windows}. Each window function $\psi(n)$ was multiplied by  $x(n)$, the full interval, without divisions in time. 

\subsection{Experimental setup}

\begin{figure}
    \begin{center}
    \begin{tabular}{c}
  \includegraphics[width=\textwidth]{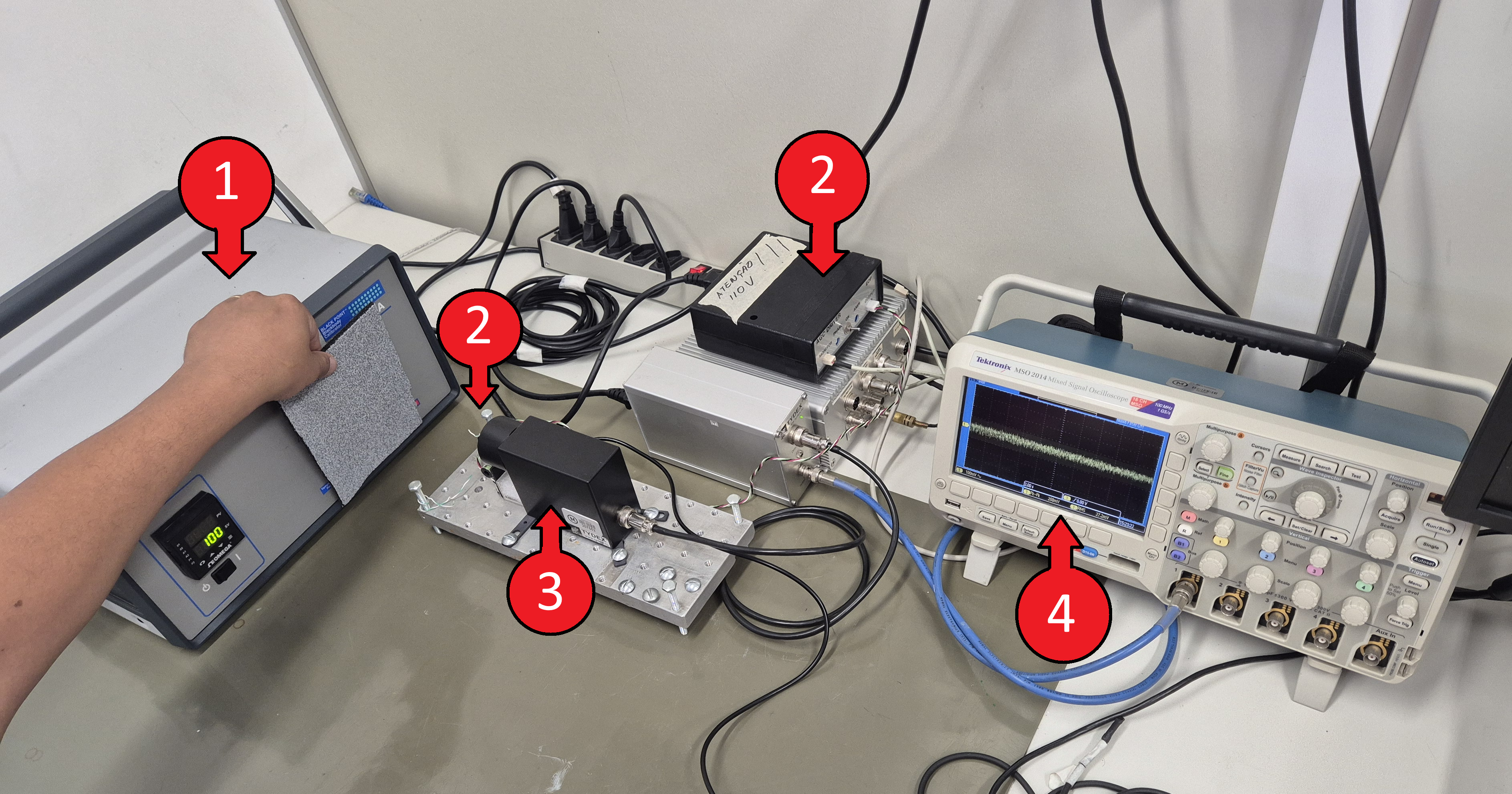}
  \end{tabular}
  \end{center}
    \caption{Bench test setup. \textbf{\textcircled{\scriptsize{1}}} Dwyer Omega BB-4A blackbody calibrator, \textbf{\textcircled{\scriptsize{2}}} Thorlabs fork chopper with MC1000A control unit, \textbf{\textcircled{\scriptsize{3}}} Tydex Golay Cell model GC-1P with amplifier unit, \textbf{\textcircled{\scriptsize{4}}} Tektronix MSO2014 digital oscilloscope. \label{fig:gerador20Hz}}
    
\end{figure}

The bench tests include equipment to test the HATS data acquisition system. The setup is shown in Figure \ref{fig:gerador20Hz}, where \textbf{\textcircled{\scriptsize{1}}} is the Dwyer Omega BB-4A blackbody calibrator, \textbf{\textcircled{\scriptsize{2}}} is the Thorlabs fork chopper with MC1000A control unit, \textbf{\textcircled{\scriptsize{3}}} is the Tydex Golay Cell model GC-1P with amplifier unit, and \textbf{\textcircled{\scriptsize{4}}} is the Tektronix MSO2014 digital oscilloscope. The fork chopper modulates the signal at 20 Hz. The Golay cell output is read with the oscilloscope. The blackbody calibrator nominally operates over the temperature range of 100 to 982 $^\circ$C. 
 
A total of 23 signals were generated, with temperatures ranging from 100 to 375 $^\circ$C (373.15 K to 648.15 K), each containing 125,000 samples, and stored in comma-separated values (CSV) files. Then we applied two different methods to obtain the calibration curve. First, we used the sine function, and then the windowed FFT.

\subsubsection{The Sine Method}

\begin{figure}
    \begin{center}
    \begin{tabular}{c}
    \includegraphics[width=0.7\textwidth]{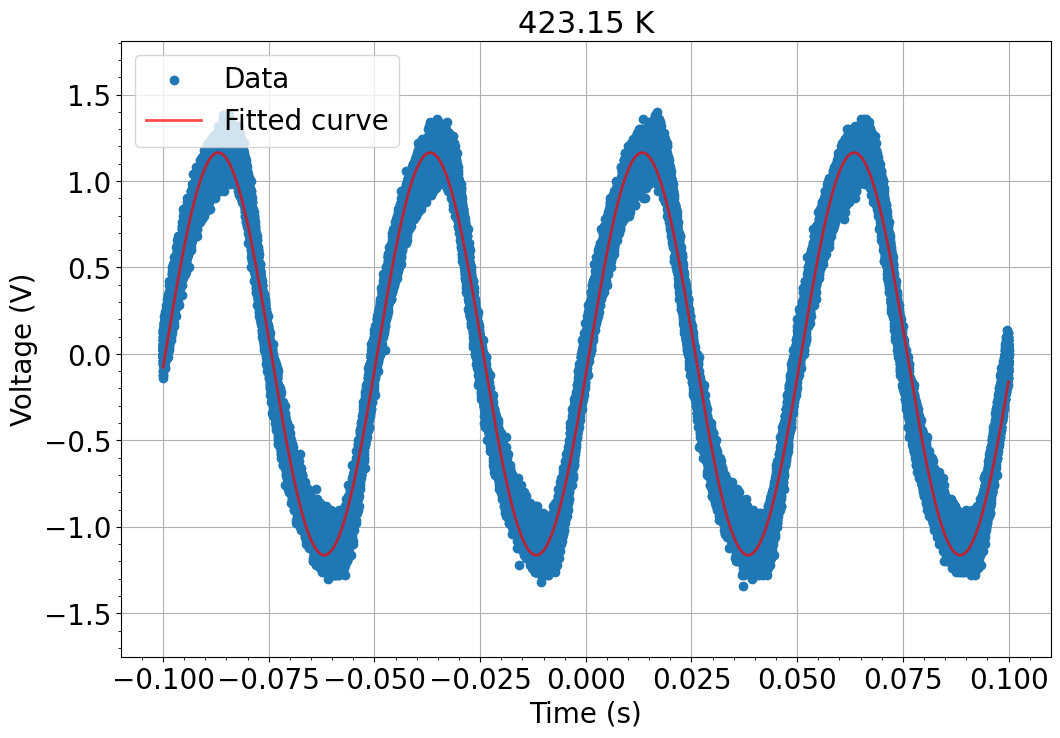}
    \end{tabular}
    \end{center}
    \caption{Registered voltage as a function of time for a source temperature of 423.15 K (blue) and the obtained fitted sine curve (red).\label{fig:voltage_time}}
    
\end{figure}

The signals were fitted to a sine function using  \texttt{curve\_fit} function, from the \texttt{scipy} library \cite{scipy_windows} and the expression 
\begin{equation}
V(t) = U \sin(\omega t + \phi),
\label{eq:sine}
\end{equation}
where $V$ is the Golay output voltage in the time domain $t$; $U$ is the sine amplitude, $\omega = 2\pi f = 125.66\ [\mathrm{s^{-1}}]$ is the angular frequency, and $\phi$ is the phase shift. Figure \ref{fig:voltage_time} shows the output voltage as a function of time (in blue), corresponding to a blackbody calibrator temperature of 
423.15~K, and the model fitting (in red). We obtained $U$ for each blackbody temperature (23 samples in total) and then fitted a linear polynomial to Eq. (\ref{eq:gain}), yielding averages for $a$ and $b$ with their standard deviations.

\subsubsection{Windowed Method}
\label{sec:calib_fft}

\begin{figure}
     \begin{center}
     \begin{tabular}{c}
     \includegraphics[height=5.5cm]{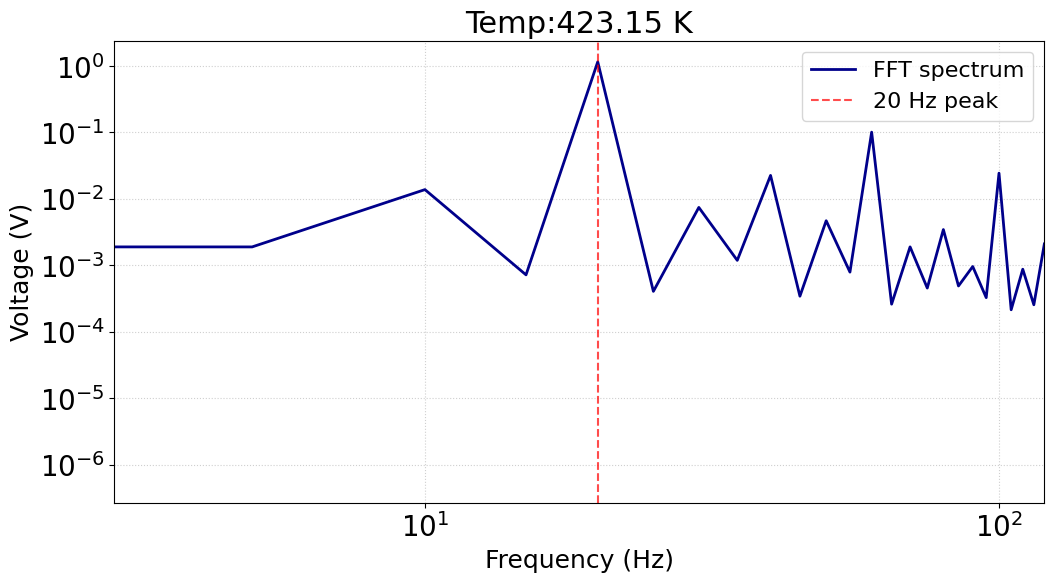}
     \end{tabular}
     \end{center}
     \caption{Signal corresponding to the temperature of 423.15 K in the frequency domain, showing the peak at 20 Hz after the FFT.\label{fig:FFT_signal}}
  
\end{figure}

To avoid signal amplitude losses in the frequency domain\cite{Harris1978,Liu2019ApplicationOF}, signals were multiplied by correction factors for each window function (see Table \ref{tab:correction_values}).   
The FFT amplitude at 20 Hz corresponds to the sine amplitude $U$. Figure \ref{fig:FFT_signal} shows the signal in the frequency domain after applying the windowed FFT. After obtaining $U$ for the 23 signal samples, we fitted a linear polynomial to the data (Eq. (\ref{eq:gain})) to get the gain $a$ and the offset $b$. 

\begin{table}[ht]
    \caption{Correction factors for different windows. \label{tab:correction_values}}
    \begin{center} 
    \begin{tabular}{|l|c|c|}
    \hline
        \textbf{Window Function} & \textbf{Correction Value} \\
        \hline
        Rectangular & 1.0 \\
        Hamming     & 1.9 \\
        Hann        & 2.0 \\
        Bartlett    & 2.0 \\
        Blackman    & 2.4 \\
        Flat-top    & 4.7 \\
        \hline
    \end{tabular}
    \end{center}
\end{table}

\section{Results}

\begin{figure}
    \begin{center}
    \begin{tabular}{c}
    \includegraphics[width=0.7\textwidth]{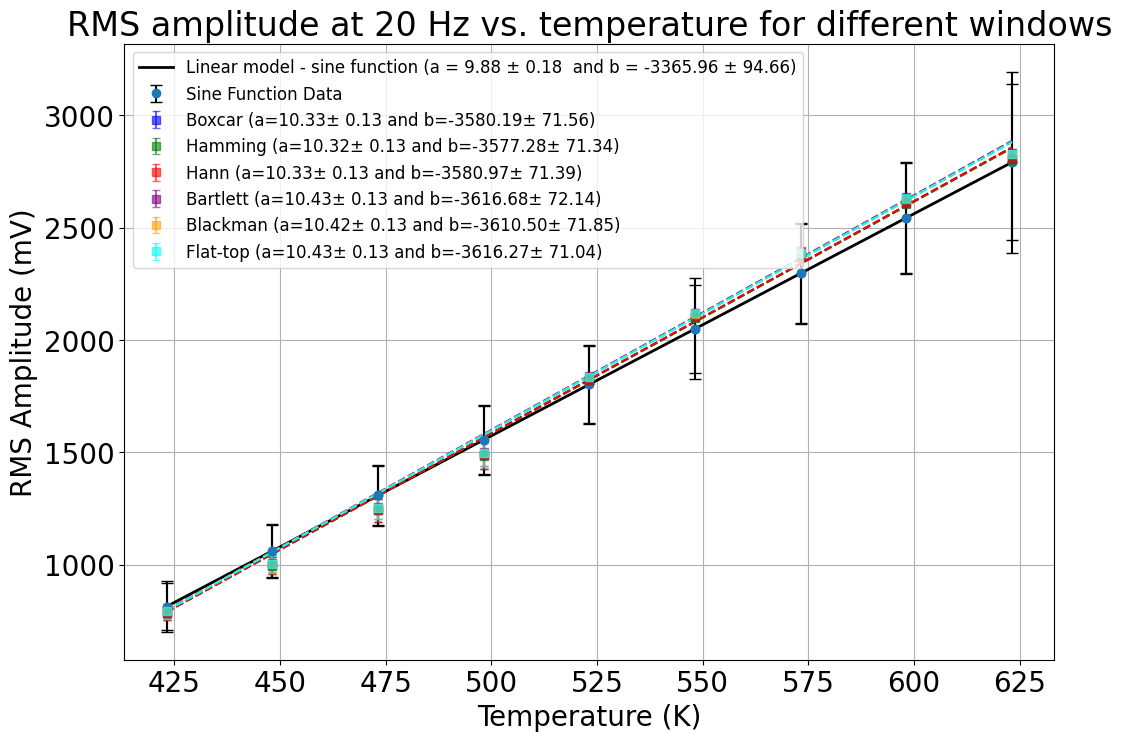}
    \end{tabular}
    \end{center}
    \caption{Comparative amplitude vs. temperature, with windowing for linear curve fitting. \label{fig:voltage_vs_temp_complete}}
    
\end{figure}

Fig. \ref{fig:voltage_vs_temp_complete} shows the fittings for the sine method (black curve) and the windowed FFT method for the different window functions. The values of the coefficients $a$ (gain) and $b$ (offset), and their uncertainties, along with the RMSE, are shown in Table \ref{table:results}. Although we present both parameters, the most important is the gain, since in astronomical observations, the background is typically subtracted from the data, and one is interested in the excess flux. As can be seen in Fig. \ref{fig:voltage_vs_temp_complete}, gain values obtained with the windowed FFT are equal within 1-$\sigma$; this fact is reinforced by the RMSE. The Hamming window yields the lowest RMSE, while the Bartlett window yields the highest. Weighted means are $a = (10.38 \pm 0.02) \ [\mathrm{mV\ K^{-1}}]$ and $b= -(3630 \pm 29) \ [\mathrm{mV}]$. The Sine Method yields $a = (9.88 \pm 0.18)  [\mathrm{mV\ K^{-1}}]$ and $b = -(3366 \pm 95) [\mathrm{mV}]$. These values differ from the weighted means of the windowed FFT method by $3\sigma$. 

\begin{table}[ht]
\caption{Linear coefficients and RMSE obtained for different window functions and the sine function fitting.\label{table:results}}
\begin{center}  
\begin{tabular}{||l|l|c|c|c||}
\hline
\textbf{Method} & & \textbf{$a$ (mV/K)} & \textbf{$b$ (mV)} & \textbf{RMSE} \\
\hline\hline
\textbf{Windowed FFT}  &&&&\\
&Rectangular & $10.33 \pm 0.13$ & $-3580 \pm 72$ & 15.62 \\ 
&Hamming              & $10.32 \pm 0.13$ & $-3577 \pm 71$ & 15.48 \\ 
&Hann                 & $10.33 \pm 0.13$ & $-3581 \pm 71$ & 15.55 \\ 
&Blackman             & $10.42 \pm 0.13$ & $-3610 \pm 72$ & 15.70 \\ 
&Bartlett             & $10.43 \pm 0.13$ & $-3617 \pm 72$ & 16.02 \\ 
&Flat-top             & $10.43 \pm 0.13$ & $-3816 \pm 71$ & 15.85 \\ 
&\textit{Weighted Mean}        & $10.38 \pm 0.02$ & $-3630 \pm 29$ & -- \\ \hline
\textbf{Sine} && $9.88 \pm 0.18$ & $-3366 \pm 95$ & -- \\ \hline\hline
\end{tabular}
\end{center}
\end{table}

\section{Conclusions}

This work presents the characterization and calibration of the ground-based HATS telescope’s data acquisition system through extensive bench tests. By utilizing a blackbody calibrator and a Golay cell, we established a precise voltage-to-temperature correlation, fundamental for converting digitized signals into physical brightness temperatures. The comparative analysis between the Sine and the windowed FFT methods demonstrates that both methodologies are highly effective for system calibration, showing strong linearity within the range of 373.15 K to 648.15 K. The application of different windowing functions proved to be a robust strategy for signal recovery. Specifically, the results indicate that although the Flat-top window presents the lowest amplitude values and the Blackman window the highest, the overall performance of all tested windows is consistent. The Hamming window yields the lowest RMSE (15.48), while the Bartlett window has the highest (16.02), suggesting that for this specific instrumentation setup, the Hamming window provides a slightly superior accuracy. However, the minimal differences between the curves confirm that the choice of window function does not significantly impact the final calibration results, provided that the appropriate correction factors are applied. Although the difference between the weighted windowed FFT means and the Sine methods is statistically significant ($3\ \sigma$) in relative terms, $a$ values differ by 5\% and $b$ values by 7.8\%. A final \textit{overall} weighted mean yields $a = (10.37\pm 0.02) \ [\mathrm{mV\ K^{-1}}]$ and $b= -(3607 \pm 28) \ [\mathrm{mV}]$. Converting $b$ to temperature yields $T_{off} = -348\ [\mathrm{K}]$. The negative sign indicates that this temperature is subtracted from the signal, and its absolute value can then be interpreted as the sensor's internal noise. 

This procedure establishes a reliable foundation for future quantitative analysis of solar flare emissions in the 15 THz range. Ongoing research will focus on the practical application of these calibration curves during solar observation campaigns to further refine the signal-to-noise ratio and enhance the detection of solar phenomena.  

\subsection*{Code, Data, and Materials Availability} 

The data and code necessary to reproduce the results in this paper are available at \\ \href{https://doi.org/10.5281/zenodo.20544685}{https://doi.org/10.5281/zenodo.20544685}. 

\subsection*{Disclosures}
The authors declare that there are no financial interests, commercial affiliations, or other potential conflicts of interest that could have influenced the objectivity of this research or the writing of this paper.

\subsection*{Acknowledgments}
We acknowledge support from Fundo Mackenzie de Pesquisa e Inovação (Mack\-Pes\-qui\-sa), Fun\-da\-ção de Amparo à Pesquisa do Estado de São Paulo (FA\-PESP) 2013\-/24155-3 and 2022\-/15700-7, Co\-or\-de\-na\-ção de Aperfeiçoamento de Pessoal de Nível Superior - Programa de Suporte à Pós-Graduação de Instituições Comunitárias de Educação Superior (CAPES - PROSUC, modality I - 2024), and from Conselho Nacional de Desenvolvimento Científico e Tecnológico (CNPq).

\bibliography{report}   

@INPROCEEDINGS{Yingxin6380076,
  author={Yingxin Wang and Ziran Zhao and Zhiqiang Chen and Linghui Wang},
  booktitle={2012 37th International Conference on Infrared, Millimeter, and Terahertz Waves}, 
  title={Characterization of Golay detector for the absolute power measurement of terahertz radiation}, 
  year={2012},
  volume={},
  number={},
  pages={1-2},
  doi={10.1109/IRMMW-THz.2012.6380076}}

@article{Golay10.1063/1.1740948,
    author = {Golay, Marcel J. E.},
    title = {Theoretical Consideration in Heat and Infra‐Red Detection, with Particular Reference to the Pneumatic Detector},
    journal = {Review of Scientific Instruments},
    volume = {18},
    number = {5},
    pages = {347-356},
    year = {1947},
    month = {05},
    issn = {0034-6748},
    doi = {10.1063/1.1740948},
    url = {https://doi.org/10.1063/1.1740948},
    eprint = {https://pubs.aip.org/aip/rsi/article-pdf/18/5/347/19136638/347_1_online.pdf},
}

@ARTICLE{Rojas2026arXiv260506344R,
       author = {{Rojas-Quesada}, Miguel and {Fletcher}, Lyndsay and {Hudson}, Hugh and {Mulay}, Sargam M. and {Simoes}, Paulo J.~A.},
        title = "{First time delay observation between two mid-infrared channels in solar flare footpoints}",
      journal = {arXiv e-prints},
         year = 2026,
        month = may,
          eid = {arXiv:2605.06344},
        pages = {arXiv:2605.06344},
          doi = {10.48550/arXiv.2605.06344},
archivePrefix = {arXiv},
       eprint = {2605.06344},
 primaryClass = {astro-ph.SR},
       adsurl = {https://ui.adsabs.harvard.edu/abs/2026arXiv260506344R}
}

@ARTICLE{Guigue2018SpWea..16.1261G,
       author = {{Gim{\'e}nez de Castro}, C.~G. and {Raulin}, J.-P. and {Valle Silva}, J.~F. and {Sim{\~o}es}, P.~J.~A. and {Kudaka}, A.~S. and {Valio}, A.},
        title = "{The 6 September 2017 X9 Super Flare Observed From Submillimeter to Mid-IR}",
      journal = {Space Weather},
         year = 2018,
        month = sep,
       volume = {16},
       number = {9},
        pages = {1261-1268},
          doi = {10.1029/2018SW001969},
       adsurl = {https://ui.adsabs.harvard.edu/abs/2018SpWea..16.1261G}
}

@ARTICLE{Lopez2022A&A...657A..51L,
       author = {{L{\'o}pez}, Fernando M. and {Gim{\'e}nez de Castro}, Carlos Guillermo and {Mandrini}, Cristina H. and {Sim{\~o}es}, Paulo J.~A. and {Cristiani}, Germ{\'a}n D. and {Gary}, Dale E. and {Francile}, Carlos and {D{\'e}moulin}, Pascal},
        title = "{A solar flare driven by thermal conduction observed in mid-infrared}",
      journal = {\aap},
         year = 2022,
        month = jan,
       volume = {657},
          eid = {A51},
        pages = {A51},
          doi = {10.1051/0004-6361/202141967},
archivePrefix = {arXiv},
       eprint = {2110.15751},
 primaryClass = {astro-ph.SR},
       adsurl = {https://ui.adsabs.harvard.edu/abs/2022A&A...657A..51L}
}

@ARTICLE{Penn2014LRSP...11....2P,
       author = {{Penn}, Matthew J.},
        title = "{Infrared Solar Physics}",
      journal = {Living Reviews in Solar Physics},
         year = 2014,
        month = dec,
       volume = {11},
       number = {1},
          eid = {2},
        pages = {2},
       adsurl = {https://ui.adsabs.harvard.edu/abs/2014LRSP...11....2P}
}

@ARTICLE{Simoes2017A&A...605A.125S,
       author = {{Sim{\~o}es}, Paulo J.~A. and {Kerr}, Graham S. and {Fletcher}, Lyndsay and {Hudson}, Hugh S. and {Gim{\'e}nez de Castro}, C. Guillermo and {Penn}, Matt},
        title = "{Formation of the thermal infrared continuum in solar flares}",
      journal = {\aap},
         year = 2017,
        month = sep,
       volume = {605},
          eid = {A125},
        pages = {A125},
          doi = {10.1051/0004-6361/201730856},
archivePrefix = {arXiv},
       eprint = {1706.09867},
 primaryClass = {astro-ph.SR},
       adsurl = {https://ui.adsabs.harvard.edu/abs/2017A&A...605A.125S}
}

@ARTICLE{Yang2025ApJ...988L..56Y,
       author = {{Yang}, Xu and {Cao}, Wenda and {Wang}, Meiqi and {Jennings}, Don and {Qiu}, Jiong and {He}, Wen and {Perriyil}, Solomon M. and {Yurchyshyn}, Vasyl and {Fletcher}, Lyndsay and {Sim{\~o}es}, Paulo J.~A. and {Jhabvala}, Murzy and {Lunsford}, Allen and {Chen}, Xingyao and {Hudson}, Hugh},
        title = "{High-resolution Observations of an X6.4 Solar Flare in the Mid-infrared}",
      journal = {\apjl},
         year = 2025,
        month = aug,
       volume = {988},
       number = {2},
          eid = {L56},
        pages = {L56},
          doi = {10.3847/2041-8213/adee95},
       adsurl = {https://ui.adsabs.harvard.edu/abs/2025ApJ...988L..56Y}
}

@ARTICLE{Fletcher2011SSRv..159...19F,
       author = {{Fletcher}, L. and {Dennis}, B.~R. and {Hudson}, H.~S. and {Krucker}, S. and {Phillips}, K. and {Veronig}, A. and {Battaglia}, M. and {Bone}, L. and {Caspi}, A. and {Chen}, Q. and {Gallagher}, P. and {Grigis}, P.~T. and {Ji}, H. and {Liu}, W. and {Milligan}, R.~O. and {Temmer}, M.},
        title = "{An Observational Overview of Solar Flares}",
      journal = {\ssr},
         year = 2011,
        month = sep,
       volume = {159},
       number = {1-4},
        pages = {19-106},
          doi = {10.1007/s11214-010-9701-8},
archivePrefix = {arXiv},
       eprint = {1109.5932},
 primaryClass = {astro-ph.SR},
       adsurl = {https://ui.adsabs.harvard.edu/abs/2011SSRv..159...19F}
}

@ARTICLE{Castro2020,
       author = {{Giménez de Castro}, C. Guillermo and {Raulin}, Jean-Pierre and {Valio}, Adriana and {Alaia}, Guilherme and {Alvarenga}, Vinicius and {Bortolucci}, Emilio Carlos and {Fernandes}, Silvia Helena and {Francile}, Carlos and {Giorgetti}, Tiago and {Kudaka}, Amauri Shossei and {López}, Fernando Marcelo and {Marcon}, Rogério and {Marun}, Adolfo and {Zaquela}, Márcio},
        title = "{HATS: A Ground-Based Telescope to Explore the THz Domain}",
      journal = {\solphys},
         year = "2020",
       volume = {295},
       number = {4},
          eid = {56},
          doi = {10.1007/s11207-020-01621-3}
}

@article{trottet2015,
  title={Origin of the 30 THz emission detected during the solar flare on 2012 March 13 at 17: 20 UT},
  author={Trottet, G and Raulin, J-P and Mackinnon, A and Gim{\'e}nez de Castro, G and Sim{\~o}es, PJA and Cabezas, D and de La Luz, V and Luoni, M and Kaufmann, P},
  journal={Solar physics},
  volume={290},
  number={10},
  pages={2809--2826},
  year={2015},
  publisher={Springer}
}

@book{stix2004,
address={New York},
author={Michael {Stix}},
edition={2},
publisher={Springer},
title={The Sun: An Introduction},
year=2004
}

@INPROCEEDINGS{fernandes2011,
  author={{Fernandes}, L. O. T. and {Kaufmann}, P. and {Marcon}, R. and {Kudaka}, A. S. and {Marun}, A. and {Godoy}, R. and {Bortolucci}, E. C. and {Zakia}, M. {Beny} and {Diniz}, J. A.},
  booktitle={2011 XXXth URSI General Assembly and Scientific Symposium}, 
  title={Photometry of THz radiation using Golay cell detector}, 
  year={2011},
  volume={},
  number={},
  pages={1-4},
  doi={10.1109/URSIGASS.2011.6051287}}

@ARTICLE{kaufmann2013,
       author = {{Kaufmann}, P. and {White}, S.~M. and {Freeland}, S.~L. and {Marcon}, R. and {Fernandes}, L.~O.~T. and {Kudaka}, A.~S. and {de Souza}, R.~V. and {Aballay}, J.~L. and {Fernandez}, G. and {Godoy}, R. and {Marun}, A. and {Valio}, A. and {Raulin}, J. -P. and {Gim{\'e}nez de Castro}, C.~G.},
        title = "{A Bright Impulsive Solar Burst Detected at 30 THz}",
      journal = {\apj},
         year = 2013,
        month = may,
       volume = {768},
       number = {2},
          eid = {134},
        pages = {134},
          doi = {10.1088/0004-637X/768/2/134},
archivePrefix = {arXiv},
       eprint = {1303.5894},
 primaryClass = {astro-ph.IM},
       adsurl = {https://ui.adsabs.harvard.edu/abs/2013ApJ...768..134K}
}

@ARTICLE{kudaka2015,
       author = {{Kudaka}, A.~S. and {Cassiano}, M.~M. and {Marcon}, R. and {Cabezas}, D.~P. and {Fernandes}, L.~O.~T. and {Hidalgo Ramirez}, R.~F. and {Kaufmann}, P. and {de Souza}, R.~V.},
        title = "{The New 30 THz Solar Telescope in S{\~a}o Paulo, Brazil}",
      journal = {\solphys},
         year = 2015,
        month = aug,
       volume = {290},
       number = {8},
        pages = {2373-2379},
          doi = {10.1007/s11207-015-0749-1},
       adsurl = {https://ui.adsabs.harvard.edu/abs/2015SoPh..290.2373K}
}

@ARTICLE{penn2016,
       author = {{Penn}, Matt and {Krucker}, S{\"a}m and {Hudson}, Hugh and {Jhabvala}, Murzy and {Jennings}, Don and {Lunsford}, Allen and {Kaufmann}, Pierre},
        title = "{Spectral and Imaging Observations of a White-light Solar Flare in the Mid-infrared}",
      journal = {\apjl},
         year = 2016,
        month = mar,
       volume = {819},
       number = {2},
          eid = {L30},
        pages = {L30},
          doi = {10.3847/2041-8205/819/2/L30},
archivePrefix = {arXiv},
       eprint = {1512.04449},
 primaryClass = {astro-ph.SR},
       adsurl = {https://ui.adsabs.harvard.edu/abs/2016ApJ...819L..30P}}

@book{prabhu2014,
address={Boca Raton},
author={K M M {Prabhu}},
edition={1},
publisher={Taylor \& Francis Group},
title={Window functions and their applications in signal processing},
year={2014}}

@ARTICLE{Bojkovic2017,
  author={{Bojkovic}, Zoran S. and {Bakmaz}, Bojan M. and {Bakmaz}, Miodrag R.},
  journal={Proceedings of the IEEE}, 
  title={Hamming Window to the Digital World}, 
  year={2017},
  volume={105},
  number={6},
  pages={1185-1190},
  doi={10.1109/JPROC.2017.2697118}}

@ARTICLE{shibata2011,
       author = {{Shibata}, Kazunari and {Magara}, Tetsuya},
        title = "{Solar Flares: Magnetohydrodynamic Processes}",
      journal = {Living Reviews in Solar Physics},
         year = 2011,
        month = dec,
       volume = {8},
       number = {1},
          eid = {6},
        pages = {6},
          doi = {10.12942/lrsp-2011-6},
       adsurl = {https://ui.adsabs.harvard.edu/abs/2011LRSP....8....6S}
}

@misc{scipy_windows,
  author       = {{SciPy Developers}},
  title        = {SciPy Signal Windows},
  year         = {2025},
  howpublished = {\url{https://docs.scipy.org/doc/scipy/reference/signal.windows.html}},
  note         = {Accessed: November 11, 2025}
}

@article{Jwo2021,
title = {Windowing Techniques, the Welch Method for Improvement of Power Spectrum Estimation},
journal = {Computers, Materials and Continua},
volume = {67},
number = {3},
pages = {3983-4003},
year = {2021},
issn = {1546-2218},
doi = {https://doi.org/10.32604/cmc.2021.014752},
url = {https://www.sciencedirect.com/science/article/pii/S152614922000168X},
author = {{Dah-Jing Jwo} and {Wei-Yeh Chang} and {I-Hua Wu}}
}

@ARTICLE{Simoes2024,
       author = {{Sim{\~o}es}, Paulo J.~A. and {Fletcher}, Lyndsay and {Hudson}, Hugh S. and {Kerr}, Graham S. and {Penn}, Matt and {Lopez}, Karla F.},
        title = "{Precise timing of solar flare footpoint sources from mid-infrared observations}",
      journal = {\mnras},
         year = 2024,
        month = jul,
       volume = {532},
       number = {1},
        pages = {705-718},
          doi = {10.1093/mnras/stae1511},
archivePrefix = {arXiv},
       eprint = {2406.11361},
 primaryClass = {astro-ph.SR},
       adsurl = {https://ui.adsabs.harvard.edu/abs/2024MNRAS.532..705S}
}

@ARTICLE{Miteva2016,
       author = {{Miteva}, R. and {Kaufmann}, P. and {Cabezas}, D.~P. and {Cassiano}, M.~M. and {Fernandes}, L.~O.~T. and {Freeland}, S.~L. and {Karlick{\'y}}, M. and {Kerdraon}, A. and {Kudaka}, A.~S. and {Luoni}, M.~L. and {Marcon}, R. and {Raulin}, J.-P. and {Trottet}, G. and {White}, S.~M.},
        title = "{Comparison of 30 THz impulsive burst time development to microwaves, H{\ensuremath{\alpha}}, EUV, and GOES soft X-rays}",
      journal = {\aap},
         year = 2016,
        month = feb,
       volume = {586},
          eid = {A91},
        pages = {A91},
          doi = {10.1051/0004-6361/201425520},
archivePrefix = {arXiv},
       eprint = {1512.01763},
 primaryClass = {astro-ph.SR},
       adsurl = {https://ui.adsabs.harvard.edu/abs/2016A&A...586A..91M}
}

@BOOK{1986rpa..book.....R,
       author = {{Rybicki}, George B. and {Lightman}, Alan P.},
        title = "{Radiative Processes in Astrophysics}",
         year = 1986,
         publisher = Wiley,
       adsurl = {https://ui.adsabs.harvard.edu/abs/1986rpa..book.....R}
}

@unpublished{Kenshima2026,
  author = {G. G. S. {Kenshima} and D. R. de {Sousa} and T. {Giorgetti} and L. A. M. {Saito} and P. J. A. {Simões} and C. G. {Giménez de Castro}} ,
  title  = {Analysis of the HATS telescope acquisition system},
  note   = {submitted for publication},
  journal = {Solar Physics},
  month  = {Jun.},
  year   = {2026}
}

@ARTICLE{Harris1978,
  author={{Harris}, F.J.},
  journal={Proceedings of the IEEE}, 
  title={On the use of windows for harmonic analysis with the discrete Fourier transform}, 
  year={1978},
  volume={66},
  number={1},
  pages={51-83},
  doi={10.1109/PROC.1978.10837}}

@article{Liu2019ApplicationOF,
  author = {{Liu}, Jian and {Yang}, Chun and {Shi}, Xiaojuan and {Wang}, Peng},
  title = {Application of FFT Interpolation Correction Algorithm Based on Window Function in Power Harmonic Analysis},
  journal = {IOP Conference Series: Earth and Environmental Science},
  volume = {252},
  pages = {032184},
  year = {2019},
  publisher = {IOP Publishing},
  doi = {10.1088/1755-1315/252/3/032184}
}

@article{GimenezdeCastroetal:2026,
   author = { {Gim{\' e}nez de Castro}, C.G. and {Francile}, C. and {L{\' o}pez}, F.M. and {Simoes} P.J.A. and {Kudaka}, A. S. and {Sousa} D.R. and {Marciani} M.G. and {Kenshima} G.S.},
   title = {HATS: a mid-IR photometer to investigate solar flares},
   year = "{in preparation}"
   }
\bibliographystyle{spiejour}   

Gedeane G. S. Kenshima is a PhD student at Mackenzie Presbyterian University.

Daniel R. Sousa is a PhD student at Mackenzie Presbyterian University.

Tiago Giorgetti is the lead engineer at Giorgetti Engenharia and a PhD student at the Polytechnic School at São Paulo University.

Paulo J. A. Simões is a professor in the Engineering School and researcher at the Center for Radio Astronomy and Astrophysics Mackenzie, at Mackenzie Presbyterian University

C. Guillermo Giménez de Castro is a professor in the Engineering School and researcher at the Center for Radio Astronomy and Astrophysics Mackenzie, at Mackenzie Presbyterian University.

\listoffigures
\listoftables

\end{spacing}
\end{document}